\documentclass[A4paper,12pt]{article}

\usepackage{amsmath}
\usepackage{a4wide}
\usepackage{amsfonts}
\newenvironment{pf}{\textbf{Proof:}}

\newtheorem{pr}{Proposition}

\title{Non Markovian Quantum Repeated Interactions and Measurements}
\author{C Pellegrini and F Petruccione\\
\vspace{-0,3cm}\scriptsize{School of Physics and National Institute for Theoretical Physics}\\
\vspace{-0,3cm}\scriptsize{University of KwaZulu-Natal}\\
\vspace{-0,3cm}\scriptsize{Private Bag X54001, Durban 4000}\\
\vspace{-0,3cm}\scriptsize{South Africa}\\
\vspace{-0,3cm}\scriptsize{e-mail: pelleg@math.univ-lyon1.fr}\\
\vspace{-0,3cm}\scriptsize{e-mail: petruccione@ukzn.ac.za}}

\begin{document}

\maketitle
\begin{abstract}
A non-Markovian model of quantum repeated interactions between a small quantum system and an infinite chain of quantum systems is presented. By adapting and applying usual projection operator techniques in this context,  discrete versions of the integro-differential and time-convolutioness Master equations for the reduced system are derived. Next, an intuitive and rigorous description of the indirect quantum measurement principle is developed and a discrete non Markovian stochastic Master equation for the open system is obtained. Finally, the question of unravelling in a particular model of non-Markovian quantum interactions is discussed.
\end{abstract}

\section{Introduction}

The theory of Open Quantum Systems describes the physical phenomena of dissipation and decoherence \cite{francesco2, davies1, Breuer3, Zurek}. Starting from the microscopic formulation of the interaction between a small system and an environment in terms of the 
\textit{Schr\"odinger equation}, there exist different ways to derive the \textit{Master equation} for the irreversible evolution of the small, i.e., reduced system. Typically, two approachs are considered: the Markovian and the non-Markovian one \cite{francesco2}. 
A very active line of research is focused on developing an appropriate description of indirect quantum measurements within both approaches. This research is motivated by recent experiments in quantum optics and quantum information \cite{haroche}.

Physically, the Markovian approach is understood as a model without memory effects of the environment In this set up the Master equation takes the form of a \textit{Lindblad-Gorini-Kossakowski-Sudarshan equation} \cite{GK, lindblad1}, where the generator of the dynamics is a completely positive map \cite{francesco2, Alicki, davies1, GK, lindblad1}. This equation is  an ordinary differential equation, which describes the evolution of the state of the small system and the study of different physical phenomena: entropy, decoherence, return to equilibrium.  Mathematically, the approach makes use of the  \textit{Born-Markov Approximation} or the \textit{Weak Coupling Limit}. Starting from the Hamiltonian description of the Schr\"odinger equation, tracing over the degrees of freedom of the environment and neglecting the memory of the interaction, the Markovian Master equation is obtained. 

In the Markovian context measurement involves a stochastic perturbation of the Lindblad equation in terms of stochastic differential equations, i.e., \textit{Stochastic Schr\"odinger equations} \cite{Barchielli2}. These equations have some  remarkable properties. First, they conserve the purity of states, i.e. unravelling (the term ``stochastic Schr\"odinger equation" is mainly used when the equation preserved purity otherwise it is called ``stochastic Master equation"). Second, the expectation of the  stochastic Schr\"odinger equation reproduces the dynamics induced by the Markovian Master equation for the density matrix of the open system. These properties are very useful for the numerical simulation of the Master equation. In fact, the so-called Monte Carlo Wave Function Method is used extensively in quantum optics and quantum information \cite{}. 

In the non-Markovian approach, memory effects of the environment give rise  to generalized Master equations, i.e.,  integro-differential equations for the density operator of the open system. Usually, these equations are obtained by projection operator techniques, e.g.  the \textit{Nakajima-Zwanzig Operator Technique} \cite{Zwa, } or the \textit{Time Convolutioness Operator Technique} \cite{francesco2}. These techniques allow for the formal description of more realistic models. However, the generalised Master equations are difficult to manipulate \cite{francesco2, francesco3}. Even if formally exact analytical solutions can be obtained, these are very hard to solve numerically.  In the non-Markovian context, the stochastic equations describing measurement procedures are usually expressed in terms of colored noise \cite{LUC, Diosi3, Diosi4} (the Markovian case involves only white noise). The justification of such models is far from being obvious and intuitive. Often, rigorous arguments are missing.  Moreover, the question of non-Markovian unravelling is still highly debated \cite{Diosi1, Diosi2, Diosi3, Diosi4, GW1, GW2, GW3, GW4, GW5}. 

In the Markovian case, a rigorous approach to the description of interaction and measurement lies in the theory of Quantum Stochastic calculus \cite{partha}. In this setup, the action of the environment (described by a Fock space) is modeled by quantum noises and the evolution is given by the solution of quantum stochastic differential equations \cite{attal, gardiner}. Quantum filtering theory \cite{bouten1, bouten2} is based on quantum stochastic calculus in order to describe quantum measurement and to derive Stochastic Schr\"odinger equations. Recently, in this spirit, a discrete model called \textit{Quantum Repeated Interactions} has been introduced \cite{attal_pautrat, attal_pautrat2}. This model provides a "useful" approximation of the interaction between a small system and an environment. The model is a small system $\mathcal{H}_0$ in contact with an infinite chain of quantum system representing the environment. All the elements of the chain are identical and independent. Pieces of the environment, denoted by $\mathcal{H}$, interact one after the other with $\mathcal{H}_0$ during a time $\tau$. Hence, by renormalizing the interaction in terms of $\tau$, quantum stochastic differential equation models can be obtained as a continuous limits ($\tau$ goes to zero) of quantum repeated interactions models. This approach has been adapted to the context of measurement in \cite{pelleg_diffusive, pelleg_poisson, pelleg_multi}. It corresponds to the model of 
\textit{Quantum Repeated Measurements}. It has been shown that stochastic 
Schr\"odinger equations can be obtained as continuous limits of the discrete version of quantum measurement. Furthermore, via concrete procedures, the approach gives an intuitive and rigorous interpretation of quantum stochastic differential equations and stochastic Schr\"odinger equations.

The main aim of this article is to present the non-Markovian model of quantum repeated interactions and discrete measurement. We define a clear mathematical model of the effect of the memory of the environment in this setup. Furthermore, we show that the natural technique (Nakajima-Zwanzig, Time convolutioness) used in the continuous non-Markovian approach can be adapted to the discrete context. We present a clear way to perform quantum repeated measurement in the non-Markovian case. As a result we obtain a rigorous discrete expression for the evolution of the small system with and without measurement. Finally, we investigate the problem of unravelling in this context. For a concrete model, we show that in the non-Markovian case unravelling  imposes a Markovian structure except for  some very special cases.
\bigskip
The article is structured as follows.

In  Section \ref{QuantumRepeated}, we present the mathematical model of quantum repeated interactions in the non-Markovian case. We adapt the presentation of \cite{attal_pautrat} to introduce memory effect of the infinite chain.

In Section \ref{DiscreteNonMarkovEvolutionEquation}, we obtain the description of the evolution of the small system by computing the Nakajima-Zwanzig and Time Convolutioness projection operator technique. Especially, we obtain a discrete version of the evolution described in the previous investigations for continuous case (see \cite{francesco2} for all details concerning the continuous version).

Section \ref{NonMarkovQuantumRepeatedInteractionsWithQuantumMeasurement} 
is devoted to the introduction of a model of measurement. We present a natural way to perform measurements in the context of non-Markovian Quantum Repeated Interactions. We define a probabilistic setup describing the random evolution of the small system. By adapting Nakajima-Zwanzig projection operator technique, we obtain a rigorous expression of an evolution equation which is a discrete version of the non-Markovian stochastic Master equation. Next, we investigate the question of unravelling by studying a special case of non-markovian quantum repeated interactions. We show that unravelling imposes strong assumptions for the evolution which in general lead to a Markovian dynamics.

\section{Quantum repeated Interaction Model}\label{QuantumRepeated}

This section is devoted to the description of the discrete model of quantum
repeated interactions in the non-Markovian setup. We start by reviewing briefly
the Markov treatment of quantum repeated
interactions \cite{attal_pautrat}. The canonical model is described as follows. A small system
$\mathcal{H}_0$ is in contact with an infinite chain of identical
and independent quantum systems (each element of the chain is
denoted by $\mathcal{H}$). Each copy of $\mathcal{H}$ interacts
with $\mathcal{H}_0$ in the following fashion. The first copy of
$\mathcal{H}$ interacts with the small system during a time $\tau$ and disappears afterwards.
Then, the second copy interacts with $\mathcal{H}_0$ during the
same time interval $\tau$ and so on. Physically, the fact that after each
interaction the copy disappears is the Markov approximation.
\smallskip

Let us now describe the mathematical setup for the non-Markovian
quantum repeated interactions. The main idea is to keep the
memory of each interaction.

As the chain is supposed to be infinite, the state space
of the chain is  described as
$$T\Phi=\bigotimes_{j=1}^\infty\mathcal{H}_j,$$
where $\mathcal{H}_j\simeq\mathcal{H}$ for all $k$. To formulate the 
precise definition of this infinite tensor product we fix
an orthonormal basis $\{X_0,\ldots,X_K\}$ of $\mathcal{H}$ (where
$K+1$ is the dimension of $\mathcal{H}$). The state $\vert
X_0\rangle\langle X_0\vert$ can be regarded as the ground state. The
basis of $T\Phi$ is constructed with respect to the stabilising sequence
induced by $X_0$.

To this end, let $\mathcal{P}$
be the set of subsets of the form $A=\{(n_1,i_1),\ldots,(n_k,i_k)\}$,
where $k\in\b{N}^\star$, 
$\{i_1,\ldots,i_k\}\in\{1,\ldots,K+1\}^k$, and $n_1<\ldots<n_k$
with $n_j\in\b{N}^\star$.
The basis of $T\Phi$ is denoted by
$\overline{B}=\{X_A,A\in\mathcal{P}\}$, where for
$A\in\mathcal{P}$ the vector $X_A$ corresponds to
$$X_A=X_0\otimes\ldots X_0\otimes X_{i_1}\otimes
X_0\otimes\ldots X_0\otimes X_{i_2}\otimes\ldots$$ and the
vector $X_{i_j}$ appears in the copy number $n_j$ of
$\mathcal{H}$.

Let us now define the basic operator. On $\mathcal{B}(\mathcal{H})$,
 the canonical operator with respect to
$\{X_0,\ldots,X_k\}$ is denoted by $a_{ij}$, that is, for all $(i,j,k)\in\{0,K\}^3$ we
have
$$a_{ij}(X_k)=\delta_{jk}\,X_i .$$
By extension, we denote by $a_{ij}^{(k)}$ the operator acting as
$a_{ij}$ on $\mathcal{H}_k$, which is the copy number $k$ of
$\mathcal{H}$. On $T\Phi$, we have
$$a_{ij}^{(k)}=I\otimes\bigotimes_{j=1}^{k-1}I\otimes a_{ij}^k\otimes\bigotimes_{j>k}I .$$
The index $k$ without round bracket means that the operator $a_{ij}$ is set on the place number $k$ is the infinite tensor product.
\smallskip

Hence, the coupled system, system and chain, is  described by the Hilbert space
\begin{equation}\label{Gamma}\Gamma=\mathcal{H}_0\otimes T\Phi. \end{equation}
We endow this Hilbert space with the following state
\begin{equation}\label{initial}\mu=\rho\otimes\bigotimes_{j=1}^{\infty}\beta_j ,\end{equation}
where $\rho$ is a reference state of $\mathcal{H}_0$ and
$\beta_j=\beta$ for all $j$, with $\beta=\vert X_0\rangle\langle X_0\vert$ is the reference state of
$\mathcal{H}$ (this corresponds to a system at zero temperature) 
\smallskip

Let us now describe the interaction setup. The first copy of
$\mathcal{H}$ interacts with $\mathcal{H}_0$ during a time
$\tau$. After this interaction a second copy of $\mathcal{H}$
interacts with $\mathcal{H}_0$ and the first copy, that is
$\mathcal{H}_2$ interacts with
$\mathcal{H}_0\otimes\mathcal{H}_1$ and so on. As a consequence,
the k-th copy of $\mathcal{H}$ interacts with
$\displaystyle{\mathcal{H}_0\otimes\bigotimes_{j=1}^{k-1}\mathcal{H}_j}$,
that is we keep the memory of the previous interactions

Mathematically, we consider a sequence of unitary operators $(U_k)$
for $k\geq1$. For each $k$ the operator $U_k$ acts non-trivially
on
$\displaystyle{\mathcal{H}_0\otimes\bigotimes_{j=1}^{k}\mathcal{H}_j}$
and acts like the identity operator on
$\displaystyle{\bigotimes_{j\geq k+1}\mathcal{H}_j}$
The sequence of unitary operators $(V_k)$ which describes
the repeated quantum interactions is defined by putting
\begin{equation}\label{V}V_k=U_k\ldots U_1\end{equation}
for all $k$ (the operator $V_k$ describes the $k$ first interactions). Hence, in the Schr\"odinger picture, after $k$ interactions, the initial
state $\mu$ defined by (\ref{initial}) becomes
\begin{equation}\label{evolution}\mu_k=V_k\mu V_k^\star.\end{equation}
It is straightforward to see that $\mu_{k+1}=U_{k+1}\mu_k
U_{k+1}^\star$ The sequence $(\mu_k)$ describes  the
evolution of the system in the quantum repeated interaction setup
with memory.
\smallskip

It is important to notice, that in the Markovian case, the unitary operator $U_k$ acts only non-trivially
on the tensor product of $\mathcal{H}_0$ with $\mathcal{H}_k$, which is the
k-th copy of $\mathcal{H}$. On the rest of the Hilbert space, it acts like the identity operator.  
In the homogeneous case fro example, the operator $U_k$ can be expressed,  for all $k$, as
\begin{equation}\label{hom}U_k=\sum_{i,j=0}^{K}U_{ij}\otimes a_{ij}^{(k)},\end{equation}
where $U_{ij}$ are operators on $\mathcal{H}_0$. In the following, to ease the notation, we suppress
the symbol $\otimes$ in similar expression 

In order to generalize the above class of Markovian models, which  have been studied extensively in \cite{attal_pautrat}, we apply the classical projection operator technique of Nakajima-Zwanzig and its time-convolutioness version to the
non-Markovian quantum repeated interactions model (see \cite{francesco2} for general introduction for Nakajima-Zwanzig and time-convolutioness operator technique). 

\section{Discrete Non Markov Evolution Equation}\label{DiscreteNonMarkovEvolutionEquation}

This section is devoted to the description of the discrete evolution of the
small system in the non-Markovian case. In the first part \ref{NZPOT}, we
apply the Nakajima-Zwanzig operator technique and in the  second
part \ref{TC}, we investigate the equivalent of time-convolutioness projection operator technique for the
discrete case.

\subsection{The Nakajima-Zwanzig Projection Operator Technique}
\label{NZPOT}

We start by applying the Nakajima-Zwanzig projection
operator technique to the sequence $(\mu_k)$ introduced in (\ref{evolution}).
For any state $\alpha$ on $\Gamma$, we define the Nakajima-Zwanzig operators:
\begin{eqnarray}\mathcal{P}\alpha&=&Tr_{T\Phi}[\alpha]\otimes\bigotimes_{j=1}^\infty\beta_j\nonumber\\
\mathcal{Q}\alpha&=&\alpha-\mathcal{P}\alpha,
\end{eqnarray}
where $Tr_{T\Phi}[\alpha]$ represents the partial trace of $\alpha$ with respect to the chain. In the canonical approach, the operator $\mathcal{P}$ projects onto the
relevant part of the small system.  The aim is to obtain an evolution equation which 
describes the sequence $(\mathcal{P}\mu_{k})$ representing
 the evolution of the relevant part of $\mathcal{H}_0$.

For a fixed $k$, the projection operators
$\mathcal{P}\mu_{k+1}$ and $\mathcal{Q}\mu_{k+1}$ are given by
\begin{eqnarray}
\label{exp}\mathcal{P}\mu_{k+1}&=&\mathcal{P}\,U_{k+1}\,\mu_k\,U_{k+1}^\star=\mathcal{P}\,U_{k+1}(\mathcal{P}
+\mathcal{Q})
\mu_k\,U_{k+1}^\star\nonumber\\&=&\mathcal{P}\,U_{k+1}\,\mathcal{P}\mu_k\,U_{k+1}^\star+
\mathcal{P}\,U_{k+1}\mathcal{Q}\mu_k\,U_{k+1}^\star ,\\\label{expr1}\mathcal{Q}\mu_{k+1}&=&\mathcal{Q}\,U_{k+1}\,\mathcal{P}\mu_k\,U_{k+1}^\star+
\mathcal{Q}\,U_{k+1\,}\mathcal{Q}\mu_k\,U_{k+1}^\star .
\end{eqnarray}
For all operators $\alpha$ on $\mathcal{B}(\Gamma)$, let denote
\begin{equation}\label{sand}\mathcal{L}_k(\alpha)=U_k\alpha U_k^\star.\end{equation}
Iterating  (\ref{expr1}) and taking into account the non-commutativity of the operator $\mathcal{L}_k$, we  obtain the following expression  for $\mathcal{Q}\mu_{k+1}$
\begin{eqnarray}\label{expr2}
\mathcal{Q}\mu_{k+1}=\sum_{i=0}^k\left(\mathcal{Q}\mathcal{L}_{k+1}\prod_{j=k}^i
\mathcal{Q}\mathcal{L}_j\,(\mathcal{P}(\mu_i))\right)+\mathcal{Q}\mathcal{L}_{k+1}\prod_{j=k}^1
\mathcal{Q}\mathcal{L}_j\,(\mathcal{Q}(\mu_0)).
\end{eqnarray}
It is important to notice, that the operator in the large round brackets in the above equation corresponds to a discrete version of the time ordering exponential term appearing in the non-Markovian continuous case (see \cite{francesco2}).
Obviously, the first term $\mathcal{Q}(\mu_0)=0$. Hence, by replacing the
expression (\ref{expr2}) in the expression (\ref{exp}), we get
\begin{eqnarray}\label{expr3}
\mathcal{P}\mu_{k+1}&=&\mathcal{P}\mathcal{L}_{k+1}\Big(\mathcal{P}\mu_k\Big)+
\mathcal{P}\mathcal{L}_{k+1}\left(\sum_{i=0}^{k-1}\mathcal{Q}\mathcal{L}_{k}\prod_{j={k-1}}^i
\mathcal{Q}\mathcal{L}_j\,\Big(\mathcal{P}\mu_i\Big)\right)\nonumber\\
&=&\mathcal{P}\mathcal{L}_{k+1}\big(\mathcal{P}\mu_k\big)+\sum_{i=0}^{k-1}\Bigg(\mathcal{P}\mathcal{L}_{k+1}\prod_{j={k}}^i
\mathcal{Q}\mathcal{L}_j\,\Big(\mathcal{P}\mu_i\Big)\Bigg) .
\end{eqnarray}

The above expression describes the complete time evolution of the relevant
part of $\mathcal{H}_0$. Like in the continuous case, it appears as a discrete integro-differential equation which involves all the history of the evolution.

It might be interesting to see how the Markovian description emerges from the more general approach described above. As in the continuous case, we obtain a discrete version of the Lindblad, Gorini, Kossakowski, Sudarshan evolution equation as is shown in the following proposition.

 \begin{pr} Let $(U_k)$ be a sequence of unitary operators which describe a Markovian evolution. For all $k$,
for all $i<k$ and for all state $\gamma$ on $\Gamma$, we have
\begin{equation}
\mathcal{P}\mathcal{L}_{k+1}\prod_{j={k}}^i
\mathcal{Q}\mathcal{L}_j\,\Big(\mathcal{P}\gamma\Big)=0 . \label{PLQLP_Prop}
\end{equation}
Hence the expression $(\ref{expr3})$ becomes
$$\mathcal{P}\mu_{k+1}=\mathcal{P}\mathcal{L}_{k+1}\Big(\mathcal{P}\mu_k\Big).$$
Furthermore, in the Markovian homogeneous case (\ref{hom}) there exists a completely positive map $\mathcal{L}$
acting on $\mathcal{B}(\mathcal{H}_0)$, such that for all $k$
\begin{eqnarray}\label{master}\mathcal{P}\mu_{k+1}=\mathcal{L}\Big(\mathcal{P}\mu_k\Big).
\end{eqnarray}
\end{pr}

\begin{pf}
Let us start by showing the last part of the proposition.
Recall
that the operator $(U_k)$, in the homogeneous Markovian case (\ref{hom}), can be expressed as
$$U_k=\sum_{ij=0}^KU_{ij}\,\,a_{ij}^{(k)}.$$
Since, for all $k$, the operator $V_k$, defined in equation (\ref{V}) acts only on the $k$ first copies of
$\mathcal{H}$, it is worth noticing that for all $k$
$$\displaystyle{\mu_k=\mathrm{Tr}_{\bigotimes_{j>k}\mathcal{H}_j}
[\mu_k]\otimes\bigotimes_{j=k+1}^\infty a_{00}^{j}}.$$ Hence for all
$k$ and all $X\in\mathcal{B}(\mathcal{H}_0)$
\begin{multline}
 \mathrm{Tr}\bigg[\mathrm{Tr}_{T\Phi}\big[\mu_{k+1}\big]\,\,X\bigg]=\hfill\\ \hphantom{ccccc}=\mathrm{Tr}\left[\mu_{k+1}\,\,X\otimes\bigotimes_{j=1}^\infty
I\right]=\mathrm{Tr}\left[\mu_{k}\,U_{k+1}^\star\left(
X\otimes\bigotimes_{j=1}^\infty I\right)U_{k+1}\right]\hfill\\
\hphantom{ccccc}=\mathrm{Tr}\bigg[\mu_k\,\,\sum_{i,j,p}U_{ij}^\star
XU_{pj}\,\, a_{ip}^{(k+1)}\bigg]\hfill\\
\hphantom{ccccc}=\mathrm{Tr}\left[\mathrm{Tr}_{\bigotimes_{j>k}\mathcal{H}_j}
[\mu_k]\otimes\bigotimes_{j=k+1}^\infty
a_{00}^j\left(\sum_{i,j,p}U_{ij}^\star
XU_{pj}\,\,a_{ip}^{(k+1)}\right)\right]\hfill\\
\hphantom{ccccc}=\mathrm{Tr}\left[\mathrm{Tr}_{\bigotimes_{m>k}\mathcal{H}_m}
[\mu_k]\otimes\bigotimes_{m>k+1}a_{00}^m\left(\sum_{j,p}U_{0j}^\star
XU_{pj}\otimes\bigotimes_{j=1}^kI\otimes
a_{0p}^{k+1}\otimes\bigotimes_{j>k+1}a_{00}^j\right)\right]\hfill\\
\hphantom{ccccc}\label{DiscreteLindblad}=\sum_{j,p}\left(\mathrm{Tr}\left[\mathrm{Tr}_{\bigotimes_{m>k}\mathcal{H}_m}
[\mu_k]\left(U_{0j}^\star
XU_{pj}\otimes\bigotimes_{j=1}^kI\right)\right]\mathrm{Tr}\left[a_{0p}^{k+1}\otimes
\bigotimes_{j>k+1}a_{00}^j\right]\right)\hfill\\
\hphantom{ccccc}=\sum_j\left(\mathrm{Tr}\left[\mathrm{Tr}_{\bigotimes_{m>k}\mathcal{H}_m}
[\mu_k]\left(U_{0j}^\star XU_{0j}\otimes\bigotimes_{j=1}^kI\right)\right]\right)\hfill\\
\hphantom{ccccc}=\mathrm{Tr}\left[\mathrm{Tr}_{T\Phi}[\mu_k]\left(\sum_jU_{0j}^\star XU_{0j}\right)\right]\hfill \\
\hphantom{ccccc}=\mathrm{Tr}\left[\left(\sum_jU_{0j}\,\Big(\mathrm{Tr}_{T\Phi}[\mu_k]\Big)\,U_{0j}^\star\right)\, X\right]\hfill\\
\hphantom{ccccc}=\mathrm{Tr}\Big[\mathcal{L}\Big(\mathrm{Tr}_{T\Phi}[\mu_k]\Big)X\Big].\hfill
\end{multline}
This proves that
$\mathcal{P}\mu_k=\mathcal{L}(\mathcal{P}\mu_k)$. It is interesting to notice, that we have obtained the Kraus decomposition of the completely positive map $\mathcal{L}$ (\ref{DiscreteLindblad}) (see \cite{attal, francesco2} for more details).  This result was obtained in a different way in \cite{attal_pautrat}. The equivalent proposition for the non homogeneous case can simply be otained by replacing the terms $U_{ij}$ by the non homogeneous terms $U_{ij}^{(k+1)}$.

Now we turn our attention to the first part of the proposition. Equation (\ref{PLQLP_Prop})
follows from the fact, that for all operators $\alpha$ of the form
\begin{equation}\label{forme}\alpha=\eta\otimes\bigotimes_{j=k+1}^{\infty} a_{00}^j,\end{equation}
with $\eta$ any operator on
$\mathcal{B}(\mathcal{H}_0\otimes\bigotimes_{j=1}^k\mathcal{H}_j)$,
we have
\begin{eqnarray}
\label{equality}
\mathcal{P}U_{k+1}\mathcal{Q}\alpha
U_{k+1}^\star=0 .
\end{eqnarray}
Indeed, with the definition of the
operation $\mathcal{P}$ and the unitary operators $(U_k)$, it is straightforward to see that
the operator $\displaystyle{\prod_{j={k}}^i
\mathcal{Q}\mathcal{L}_j\,\Big(\mathcal{P}\gamma}\Big)$ is of the
same form as an operator  $\alpha$  (\ref{forme}).

Now, we are in the position to prove the result (\ref{equality}). The operator $\alpha$
can be expressed as
$$\alpha=\sum_{(i_1,j_1),\ldots,(i_k,j_k)\in\{0,\ldots,K\}}
\alpha_{(i_1,j_1),\ldots,(i_k,j_k)}\otimes a^1_{i_1,j_1}
\otimes\ldots\otimes
a^k_{i_k,j_k}\otimes\bigotimes_{j>k}a^j_{00}.$$  Then, with our notation and with the rule of the partial trace, one can see that
$$\mathcal{P}\alpha=\sum_{i_1,\ldots,i_k}\alpha_{(i_1,i_1),\ldots,(i_k,i_k)}\otimes
\bigotimes_{j=1}^\infty a_{00}^j $$
 Accordingly, for $\mathcal{Q}\alpha$,
we obtain
\begin{equation}
\mathcal{Q}\alpha=\sum_{(i_1,j_1),\ldots,(i_k,j_k)\in\{0,\ldots,K\}}
\beta_{(i_1,j_1),\ldots,(i_k,j_k)}\otimes a^1_{i_1,j_1}
\otimes\ldots\otimes
a^k_{i_k,j_k}\otimes\bigotimes_{j>k}a^j_{00}, \label{Qalpha}
\end{equation} 
with
\begin{eqnarray*}
\hphantom{ccccccc}\beta_{(0,0),\ldots,(0,0)}&=&-\sum_{(i_1,\ldots,i_k)
\neq(0,\ldots,0)}\alpha_{(i_1,i_1),\ldots,(i_k,i_k)}\\
\hphantom{ccccccc}\beta_{(i_1,j_1),\ldots,(i_k,j_k)}&=&\alpha_{(i_1,j_1),\ldots,(i_k,j_k)},\,\,
\textrm{for
all}\,\,(i_1,j_1),\ldots(i_k,j_k)\neq(0,0),\ldots,(0,0).
\end{eqnarray*}
By applying $\mathcal{L}_{k+1}$ on $\mathcal{Q}\alpha$ (\ref{Qalpha}), we get
\begin{eqnarray*}
 U_{k+1}\mathcal{Q}\alpha U_{k+1}^\star = \\
\hphantom{cc}=\sum_{ij}
\sum_{(i_1,j_1),\ldots,(i_k,j_k)\in\{0,\ldots,K\}}U_{i0}\,\beta_{(i_1,j_1),\ldots,(i_k,j_k)}\,
U_{j0}^\star\otimes a^1_{i_1,j_1} \otimes\ldots\otimes
a^k_{i_k,j_k}\otimes a_{ij}\otimes\bigotimes_{j>k+1}a^j_{00} 
\end{eqnarray*}
The result (\ref{equality}) is proved by noticing that
\begin{eqnarray*}\mathcal{P}U_{k+1}\mathcal{Q}\alpha U_{k+1}^\star&=&\sum_{i}
\sum_{i_1,\ldots,i_k\in\{0,\ldots,K\}}U_{i0}\,\beta_{(i_1,i_1),\ldots,(i_k,i_k)}\,
U_{i0}^\star\otimes\bigotimes_{j=1}^\infty a^j_{00}\\
\hphantom{\mathcal{P}U_{k+1}\mathcal{Q}\alpha U_{k+1}^\star}&=&\sum_{i}U_{i0}\,\beta_{(0,0)\ldots(0,0)}\,U_{i0}^\star\otimes\bigotimes_{j=1}^\infty
a^j_{00}\\&&+\sum_{i}U_{i0}
\sum_{i_1,\ldots,i_k\neq0,\ldots,0}\,\alpha_{(i_1,i_1),\ldots,(i_k,i_k)}\,
U_{i0}^\star\otimes\bigotimes_{j=1}^\infty a^j_{00}\\&=&0.
\end{eqnarray*}
This completes the proof of the proposition.
\end{pf}

\subsection{The Time-convolutioness Projection Operator Method}\label{TC}

In this Section, we apply the time convolutioness version of the Nakajima-Zwanzig operator technique to the framework of non-Markovian quantum repeated interactions. The aim is to derive a discrete time-evolution equation for $(\mathcal{P}\mu_{k})$ which links $\mathcal{P}\mu_{k+1}$ only to $\mathcal{P}\mu_{k}$.
 
To this end, for all $p$, we define the inverse of the operator $\mathcal{L}_p$ as
$$\mathcal{L}_p^{(-1)}(\alpha)=\big(U_p^{-1})\alpha\big(U_p^{-1})^\star,$$  for all operators $\alpha$.
It is straightforward  to see that for all $p$ we have $\mu_p=\mathcal{L}_{p+1}^{(-1)}(\mu_{p+1})$. Hence, the expression (\ref{expr2}) obtained in the previous section becomes
\begin{eqnarray}\label{expr5}
\mathcal{Q}\mu_{k+1}&=&\sum_{i=0}^k\mathcal{Q}\mathcal{L}_{k+1}\prod_{j=k}^i
\mathcal{Q}\mathcal{L}_j\,\mathcal{P}\prod_{p=i+1}^{k+1}\mathcal{L}_p^{(-1)}(\mu_{k+1})\nonumber\\
&=&\sum_{i=0}^k\mathcal{Q}\mathcal{L}_{k+1}\prod_{j=k}^i
\mathcal{Q}\mathcal{L}_j\,\mathcal{P}\prod_{p=i+1}^{k+1}\mathcal{L}_p^{(-1)}\big(\mathcal{P}+
\mathcal{Q} \big)(\mu_{k+1})\nonumber\\
&=&\sum_{i=0}^k\mathcal{Q}\mathcal{L}_{k+1}\prod_{j=k}^i
\mathcal{Q}\mathcal{L}_j\,\mathcal{P}\prod_{p=i+1}^{k+1}\mathcal{L}_p^{(-1)}\Big(\mathcal{P}
\mu_{k+1}\Big)\nonumber\\&&+\sum_{i=0}^k\mathcal{Q}\mathcal{L}_{k+1}\prod_{j=k}^i
\mathcal{Q}\mathcal{L}_j\,\mathcal{P}\prod_{p=i+1}^{k+1}\mathcal{L}_p^{(-1)}\Big(\mathcal{Q}
\mu_{k+1}\Big)\nonumber\\
&=&\sum_{i=0}^k\Bigg(\prod_{j=k+1}^{i-1}
\mathcal{Q}\mathcal{L}_j\Bigg)\mathcal{Q}\mathcal{L}_i\,\mathcal{P}
\Bigg(\prod_{p=i+1}^{k+1}\mathcal{L}_p^{(-1)}\Bigg)\Big(\mathcal{P}
\mu_{k+1}\Big)\nonumber\\&&+\sum_{i=0}^k\Bigg(\prod_{j=k+1}^{i-1}
\mathcal{Q}\mathcal{L}_j\Bigg)\mathcal{Q}\mathcal{L}_i\,\mathcal{P}\Bigg(\prod_{p=i+1}^{k+1}
\mathcal{L}_p^{(-1)}\Bigg)\Big(\mathcal{Q} \mu_{k+1}\Big).
\end{eqnarray}
For a fixed $m$, we define for all operators $\alpha\in\mathcal{B}(\Gamma)$
the following 
$$\mathcal{K}_{m+1}(\alpha)=\sum_{i=0}^m\Bigg(\prod_{j=m+1}^{i-1}
\mathcal{Q}\mathcal{L}_j\Bigg)\mathcal{Q}\mathcal{L}_i\,\mathcal{P}\Bigg(\prod_{p=i+1}^{m+1}
\mathcal{L}_p^{(-1)}\Bigg)\big(\alpha\big).$$ The expression
(\ref{expr5}) becomes
\begin{eqnarray}
\bigg[I-\mathcal{K}_{k+1}\bigg]\mathcal{Q}\mu_{k+1}=\mathcal{K}_{k+1}\mathcal{P}\mu_{k+1}.
\end{eqnarray}
Hence, if we assume that $\bigg[I-\mathcal{K}_{k+1}\bigg]$ is
invertible, we get
\begin{eqnarray}
\mathcal{Q}(\mu_{k+1})=\bigg[I-\mathcal{K}_{k+1}\bigg]^{(-1)}
\mathcal{K}_{k+1}\mathcal{P}\mu_{k+1}.
\end{eqnarray}
As a consequence, for the expression of $\mathcal{P}\mu_{k+1}$ can now be written as
\begin{eqnarray}\label{expr6}
\mathcal{P}\mu_{k+1}
&=&\mathcal{P}\mathcal{L}_{k+1}\mathcal{P}\mu_k+\mathcal{P}\mathcal{L}_{k+1}\Bigg[\bigg[I-\mathcal{K}_{k}\bigg]^{(-1)}
\mathcal{K}_{k}\mathcal{P}\mu_{k}\Bigg].
\end{eqnarray}
The expression $(\ref{expr6})$ is the discrete time-convolutionless equation of
evolution for the reduced system. The invertibility of $I-\mathcal{K}_{k}$ implies  the discrete equivalent of the locality in the time continuous case (see \cite{francesco2} for more details).

\section{Non Markov Quantum Repeated Interactions With Quantum
Measurement}
\label{NonMarkovQuantumRepeatedInteractionsWithQuantumMeasurement}

In the Markovian case the theory of quantum measurement and the corresponding evolution are well understood and well studied. This is not the case in a  non-Markovian set up. One of the open questions concerns the description of the evolution of the reduced system undergoing indirect quantum measurement.  The way to describe the random non-Markovian evolution of the reduced system in the continuous time case is far from obvious and in general it is based on heavy technique \cite{Diosi4, GW4}.

In this section, we present a very clear mathematical and physical
way to describe indirect quantum measurement in the discrete non-Markovian
setup. The approach consists of adapting the quantum repeated measurement \cite{attal_pautrat}
in the non-Markovian quantum repeated interactions model. We will show how to derive  a discrete non Markovian stochastic Master equation.

We proceed in the following way. After each
interaction, implying a new copy of $\mathcal{H}$, a quantum
measurement is performed on the last copy which has interacted.
Each measurement involves a random evolution of the state of the
system $\Gamma$. The repeated sequence of measurement gives  rise
to a random sequence of states on $\Gamma$. Then, we apply the Nakajima-Zwanzig operator technique on this sequence in order to obtain the evolution of the relevant part of the
reduced system $\mathcal{H}_0$. In fact, this strategy is very natural and does not require any phenomenological inputs.

After this, we will consider the  problem of unravelling.
Obtaining unravelling in the non-Markovian case imposes
strong assumptions on the interaction setup. Focusing on a special case, we show that essentially unravelling implies the Markov assumption except for very special situations.

\subsection{Quantum Repeated Measurements}

We start by describing the setup of quantum repeated
measurements on the whole system.  After each interaction, a measurement of an
observable is performed on the last copy of $\mathcal{H}$ which
has interacted.

We need to introduce some notations. Let $A=\sum_{i=0}^p\lambda_i P_i$ be the spectral decomposition of an
observable of $\mathcal{H}$. In analogy to the construction of the basic operators $a_{ij}^{(k)}$ in Section 2, we introduce 
$$A^{(k)}=I\otimes\bigotimes_{j=1}^{k-1}I\otimes A\otimes\bigotimes_{j>k+1}I.$$
The operator $A^{(k)}$, which  is an observable of $\mathcal{H}_k$, is the extension of the observable 
$A$ to the whole space $\Gamma$ (\ref{Gamma}). In the same way, for all $i\in\{0, \ldots, p\}$ we
denote the eigen-projectors as
$$P_i^{(k)}=I\otimes\bigotimes_{j=1}^{k-1}I\otimes P_i\otimes\bigotimes_{j>k+1}I.$$ 

Let us describe a single interaction and a single measurement.
After the first  interaction the state of the system is described by
$\mu_1=U_1\mu U_1^\star$, where $\mu$ is the initial state (\ref{initial}). 
According to
the postulates of quantum 
measurement only the eigenvalues of $A^{(1)}$
can be observed; the result is random and obeys to the
probability law
$$P[\textrm{to observe}\,\,\lambda_i]=\mathrm{Tr}\left[\,\mu_1\, P^{(1)}_i\,\right].$$
Furthermore, if we have observed the eigenvalue $\lambda_i$, the
state of the system becomes
$$\rho_1(i)=\frac{P^{(1)}_i\,\mu_1\,P^{(1)}}{\mathrm{Tr}\left[\,\mu_1\, P^{(1)}_i\,\right]}.$$
The new state $\rho_1$ is actually a random variable.
It describes the result of one
interaction and one measurement.

Let us make precise the probabilistic framework of the complete procedure of quantum repeated
measurements. To this end, we introduce the probabilistic space  
$$\Sigma=\{0,1,\ldots,p\}^{\b{N}^\star},$$ 
where the index $i$ corresponding to the eigenvalue $\lambda_i$. We endow
this space with the cylinder algebra $\mathcal{C}$ generated by
the set
$$\Lambda_{i_1,\ldots,i_k}=\left\{\omega\in\Sigma\mid\omega_1=i_1,\ldots,\omega_k=i_k\right\}.$$
Now, we define a probability law on the cylinder set. 
To this end, we introduce for all $(i_1,\ldots,i_k)\in\{0,\ldots,p\}^k:$ 
\begin{eqnarray}
\tilde{\mu}_k(i_1,\ldots,i_k)&=&\prod_{j=k}^1P^{(i)}_{i_j}U_i\,\big(\mu\big)\,\prod_{j=1}^kU_i^\star
P^{(i)}_{i_j}.
\end{eqnarray}
To simplify the above expression, we introduce the notation
 $$\overline{P}^{(i)}_{i_j}[\alpha]=P^{(i)}_{i_j}\,\,\alpha\,\,
 P^{(i)}_{i_j}.$$
Hence, by using the definition of the operations $\mathcal{L}_k$, we get
\begin{eqnarray}
\tilde{\mu}_k(i_1,\ldots,i_k)&=&\prod_{j=k}^1\overline{P}^{(i)}_{i_j}\mathcal{L}_i\,\big(\mu\big).
\end{eqnarray}
This corresponds to the non-normalized state, if we have observed
the eigenvalues $\lambda_{i_1},\ldots,\lambda_{i_k}$. The probability law on the cylinder set is now defined as
$$P[\Lambda_{i_1,\ldots,i_k}]=\mathrm{Tr}\Big[\tilde{\mu}_k
(i_1,\ldots,i_k)\Big].$$
 It is easy to check that the above definition satisfies the Kolmogorov
 consistency criterion. As a consequence, it defines a unique probability law $P$ on
$(\Sigma,\mathcal{C})$. Now
 for all $k$ and all $\omega\in\Sigma$ we can define the following random variable
$$\rho_k(\omega)=\frac{\tilde{\mu}_k(\omega_1,\ldots,\omega_k)}{\mathrm{Tr}\Big[\tilde{\mu}_k
(\omega_1,\ldots,\omega_k)\Big]}.$$ The
quantum repeated interactions combined with quantum repeated
measurements are then described by the random sequence $(\rho_k)$.
Such a sequence is called a \textit{discrete quantum
trajectory} (on the whole space). To complete the description of discrete quantum trajectories on the whole space, we have the following Markov property.

\begin{pr}\label{mark}
The random sequence $(\rho_k)$ is a Markov chain on
$(\Omega,\mathcal{C},P)$ valued in the set of states on $\Gamma$.

Furthermore, if $\rho_k=\theta$ is a state on $\Gamma$ the random
variable $\rho_{k+1}$ takes the values
$$\rho_{k+1}=\frac{\overline{P}^{(k+1)}_{i}\mathcal{L}_{k+1}(\rho_k)}
{\mathrm{Tr}\left[\overline{P}^{(k+1)}_{i}\mathcal{L}_{k+1}(\rho_k)\right]},\,\,i=0,\ldots,p,$$
with probability
$p_i=\mathrm{Tr}\left[\overline{P}^{(k+1)}_{i}\mathcal{L}_{k+1}(\rho_k)\right]$.
\end{pr}

In the next section, we apply the Nakajima-Zwanzig projection operator technique to the random sequence $(\rho_k)$. As in the case without measurement (see Section \ref{NZPOT}), this  allows for the description of the evolution of the relevant part of the system. The projection on the reduced system gives rise to a random sequence of states of $\mathcal{H}_0$, which is, in general, non-Markovian.

\subsection{Non-Markovian Stochastic Discrete Evolution Equation}

This section is devoted to the description of the discrete stochastic evolution equation  of the reduced system obtained by the Nakajima-Zwanzig projection operator technique. The equation we obtain is a discrete version of the non-Markovian stochastic Master equations (see \cite{Diosi4, GW4} for continuous version).

Let $(\rho_k)$ be a quantum trajectory as described in Proposition
\ref{mark}. For all $k$ and for all $i\in\{0,\ldots,p\}$ we denote the transition probabilities by
$p_i^{k+1}=\mathrm{Tr}[\overline{P}^{(k+1)}_{i}\mathcal{L}_{k+1}(\rho_k)]$.
Hence, with Proposition $\ref{mark}$, we can describe the evolution of $(\rho_k)$ by the following equation
\begin{eqnarray}\label{exprtotal}
\rho_{k+1}&=&\sum_{i=0}^p\frac{\big(\overline{P}^{k+1}_i
\mathcal{L}_{k+1}\big)(\rho_k)} {p^{k+1}_i}\,\mathbf{1}_i^{k+1},
\end{eqnarray}
where the indicator functions $\mathbf{1}_i^{k+1}$ corresponds to the observation of the eigenvalues $\lambda_i$ during the $(k+1)$-th measurement.  By applying the Nakajima-Zwanzig operator $\mathcal{P}$ and $\mathcal{Q}$ introduced in Section \ref{NZPOT} to (\ref{exprtotal}), we get
\begin{eqnarray*}
\mathcal{P}\rho_{k+1}&=&\mathcal{P}\left(\sum_{i=0}^p\frac{\big(\overline{P}^{k+1}_i
\mathcal{L}_{k+1}\big)(\rho_k)} {p^{k+1}_i}
\,\mathbf{1}_i^{k+1}\right)\\&=&
\sum_{i=0}^p\frac{1}{p^{k+1}_i}\mathcal{P}\big(\overline{P}^{k+1}_i\mathcal{L}_{k+1}\big)
(\mathcal{P}\rho_k)\,\mathbf{1}_i^{k+1}+\sum_{i=0}^p\frac{1}{p^{k+1}_i}
\mathcal{P}\big(\overline{P}^{k+1}_i\mathcal{L}_{k+1}\big)
(\mathcal{Q}\rho_k)\,\mathbf{1}_i^{k+1},\\
\mathcal{Q}\rho_{k+1}&=&\mathcal{Q}\left(\sum_{i=0}^p\frac{\big(\overline{P}^{k+1}_i
\mathcal{L}_{k+1}\big)(\rho_k)} {p^{k+1}_i}
\,\mathbf{1}_i^{k+1}\right)\\&=&
\sum_{i=0}^p\frac{1}{p^{k+1}_i}\mathcal{Q}\big(\overline{P}^{k+1}_i\mathcal{L}_{k+1}\big)
(\mathcal{P}\rho_k)\,\mathbf{1}_i^{k+1}+\sum_{i=0}^p\frac{1}{p^{k+1}_i}\mathcal{Q}
\big(\overline{P}^{k+1}_i\mathcal{L}_{k+1}\big)
(\mathcal{Q}\rho_k)\,\mathbf{1}_i^{k+1}.
\end{eqnarray*}
By iteration, as is done in the case without measurement, we get the following expression for
$\mathcal{Q}\rho_{k+1}$
\begin{eqnarray}
\mathcal{Q}\rho_{k+1} &=&\sum_{j=0}^{k}\sum_{(i_{k+1},\ldots,
i_j)\in\{0,\ldots,p\}}\prod_{l=k+1}^{j}\mathcal{Q}\big(\overline{P}^{l}_{i_l}\mathcal{L}_{l}\big)
(\mathcal{P}\rho_{j})\prod_{l=k+1}^{j}
\frac{\mathbf{1}^{l}_{i_l}}{p^{l}_{i_l}}\nonumber\\&&+
\sum_{(i_{k+1},\ldots,
i_1)\in\{0,\ldots,p\}}\prod_{l=k+1}^{1}\mathcal{Q}\big(\overline{P}^{l}_{i_l}\mathcal{L}_{l}\big)
(\mathcal{Q}\rho_{0})\prod_{l=k+1}^{1}
\frac{\mathbf{1}^{l}_{i_l}}{p^{l}_{i_l}}\nonumber\\
&=&\sum_{j=0}^{k}\sum_{(i_{k+1},\ldots,
i_p)\in\{0,\ldots,p\}}\prod_{l=k+1}^{j}\mathcal{Q}\big(\overline{P}^{l}_{i_l}\mathcal{L}_{l}\big)
(\mathcal{P}\rho_{j})\prod_{l=k+1}^{j}
\frac{\mathbf{1}^{l}_{i_l}}{p^{l}_{i_l}}. \label{QRhokp1}
\end{eqnarray}
The last equality in (\ref{QRhokp1}) comes from the fact that
$\mathcal{Q}(\rho_0)=\mathcal{Q}(\mu_0)=0$. Hence, we get the
following equation
\begin{eqnarray}\label{non-mark}
\mathcal{P}\rho_{k+1}&=&\sum_{i=0}^p\frac{1}{p^{k+1}_i}\mathcal{P}\big(\overline{P}^{k+1}_i
\mathcal{L}_{k+1}\big)(\mathcal{P}\rho_k)
\,\mathbf{1}_i^{k+1}\nonumber\\
&&+\sum_{j=0}^{k-1}\sum_{(i_{k+1},i_k,\ldots,
i_1)\in\{0,\ldots,p\}}\mathcal{P}\big(\overline{P}^{k+1}_{i_{k+1}}\mathcal{L}_{k+1}\big)
\left(\prod_{l=k}^{j}\mathcal{Q}\big(\overline{P}^{l}_{i_l}\mathcal{L}_{l}\big)
(\mathcal{P}\rho_{j})\right)\prod_{l=k+1}^{j}
\frac{\mathbf{1}^{l}_{i_l}}{p^{l}_{i_l}}.
\end{eqnarray}
In order to express the above equation only with terms
$\mathcal{P}\mu_i$, it is worth noticing that for all $k$ and for
all $i$
\begin{eqnarray*}
p_i^{k+1}&=&\mathrm{Tr}\left[\overline{P}^{(k+1)}_{i}\mathcal{L}_{k+1}(\rho_k)\right]\\
&=&\mathrm{Tr}\left[\mathcal{P}\left(\overline{P}^{(k+1)}_{i}\mathcal{L}_{k+1}(\rho_k)\right)\right]\\
&=&\mathrm{Tr}\left[\mathcal{P}\left(\overline{P}^{(k+1)}_{i}\mathcal{L}_{k+1}
(\mathcal{P}\rho_k)\right)\right]+\mathrm{Tr}\left[\mathcal{P}
\left(\overline{P}^{(k+1)}_{i}\mathcal{L}_{k+1}(\mathcal{Q}\rho_k)\right)\right].
\end{eqnarray*}
By replacing the expression of $\mathcal{Q}\rho_k$ with (\ref{QRhokp1}),
we get an expression involving only terms $\mathcal{P}\mu_i$.

The expression (\ref{non-mark}) describes then the random
evolution of the relevant part of the system. It is
typically a non-Markovian chain evolution, because the expression at
time $k$ involves all the past of the sequence. The equation (\ref{non-mark}) is the discrete version of the non-Markovian stochastic Master equation. The sequence $(\mathcal{P}\mu_k)$ is called the \textit{reduced quantum trajectory}.

In the Markovian case (see \cite{pelleg_multi} for a complete study), by applying a result similar to Proposition \ref{mark}, the evolution equation (\ref{non-mark}) is reduced to
$$\mathcal{P}\rho_{k+1}=\sum_{i=0}^p\frac{1}{p^{k+1}_i}\mathcal{P}\big(\overline{P}^{k+1}_i
\mathcal{L}_{k+1}\big)(\mathcal{P}\rho_k) \,\mathbf{1}_i^{k+1}$$
and describes a Markov chain. The last subsection is devoted to the investigation of the pure state
evolution in the non-Markovian case.

\subsection{Non Pure State Quantum Trajectory}

The success of the Indirect Quantum Measurement Theory and the
Quantum Trajectory Theory in the Markovian case lies in the fact
that the indirect measurement gives rise to a pure state
evolution for the reduced system. This means that, if initially
the state of $\mathcal{H}_0$ is pure, the reduced quantum trajectory
evolves on the set of pure states of $\mathcal{H}_0$ \cite{barchielli1, francesco2}. As is made precise in the Introduction, such a property is widely used in Monte Carlo Wave Function methods  for the simulation of the Quantum Markov Master Equation. 

In this section, we start by recalling the pure
state property for the discrete quantum trajectory in the
Markovian case. Next, for a specific model, we investigate the conditions under which a pure state trajectory can be unravelled in the non-Markovian case. 

\subsubsection{Markovian unravelling}\label{tt}

 In this subsection, we show that the measurement of an observable which has different eigenvalues (with eigen-space of dimension 1) provides the unravelling result in the Markovian case.
\begin{pr}
Let $(U_k)$ be a sequence of unitary operators describing a
quantum repeated interaction setup in the Markovian case. Let $A$
be an observable whose spectral decomposition is given by
$$A=\sum_{i=0}^{K}\lambda_iP_i,$$
where $K+1$ is the dimension of $\mathcal{H}$. Let
$\rho_0=\vert\psi_0\rangle\langle\psi_0\vert$ be the initial
state of $\mathcal{H}_0$. Let $(\mathcal{P}\rho_k)$ be the reduced quantum
trajectory describing the evolution of the state of
$\mathcal{H}_0$ in the setup of quantum repeated measurements of
$A$.

Then, there exists a random sequence of vectors $(\psi_k)$ of
$\mathcal{H}_0$ such that
\begin{equation}\label{pur}\mathcal{P}\rho_k=\vert\psi_k\rangle\langle\psi_k\vert\otimes\bigotimes_{j=1}^\infty\vert X_0
\rangle\langle X_0\vert.\end{equation}
\end{pr}
\begin{pf}
By recursion, suppose that there exists $\psi_k$ such that
$\mathcal{P}\rho_k=\vert\psi_k\rangle\langle\psi_k\vert
\otimes\bigotimes_{j=1}^\infty\vert X_0\rangle\langle X_0\vert$.
As the evolution of $\mathcal{P}(\rho_k)$ in the Markovian case is given by
$$\mathcal{P}\rho_{k+1}=\sum_{m=0}^K\frac{1}{p^{k+1}_m}\mathcal{P}\big(\overline{P}^{k+1}_m
\mathcal{L}_{k+1}\big)(\mathcal{P}\rho_k) \,\mathbf{1}_m^{k+1},$$
it is sufficient to show that each $\overline{P}^{k+1}_m
\mathcal{L}_{k+1}\big(\mathcal{P}\rho_k)/p_m^{k+1}$ can be expressed as (\ref{pur}). To this end, we make explicit the expression of $\overline{P}^{k+1}_m
\mathcal{L}_{k+1}\big(\mathcal{P}\rho_k)$ for all $m\in\{0,\ldots,K\}$. Let us consider a homogeneous Markovian evolution (the non homogeneous case can be easily adapted). With the same
notation as in the proof of Proposition \ref{mark}, we get the following for $\mathcal{L}_{k+1}\big(\mathcal{P}\rho_k)$
$$\mathcal{L}_{k+1}\big(\mathcal{P}\rho_k)=\sum_{ij}U_{i0}\vert\psi_k\rangle\langle\psi_k\vert
U_{j0}^\star\otimes\bigotimes_{l=1}^{k}a_{00}^l\otimes
a_{ij}^{k+1}\otimes\bigotimes_{l=k+2}^\infty a_{00}^l.$$  As the observable $A$ has eigenvalues of rank one, for all
$m\in\{0,\ldots,K\}$ the
projector $P_m$ is a rank one projector. Hence there exists a
unitary operator $Q_m$ such that $P_m=\vert Q_m X_0\rangle\langle
Q_m X_0\vert $. This allows to write \begin{eqnarray*}&&\overline{P}^{k+1}_m
\mathcal{L}_{k+1}\big(\mathcal{P}\rho_k)=\\&=&\sum_{ij}U_{i0}\vert\psi_k\rangle\langle\psi_k\vert
U_{j0}^\star\otimes\bigotimes_{l=1}^{k}a_{00}^l\otimes \langle
Q_mX_0,X_i\rangle\langle X_j,Q_mX_0\rangle\vert
Q_mX_0\rangle\langle
Q_mX_0\vert \otimes\bigotimes_{l=k+2}^\infty a_{00}^l\\&=&
\bigg\vert \sum_i \langle Q_mX_0,X_i\rangle
U_{i0}\psi_k\bigg\rangle\bigg\langle\sum_j \langle
Q_mX_0,X_j\rangle
U_{j0}\psi_k\bigg\vert\otimes\bigotimes_{l=1}^{k}a_{00}^l\otimes\vert
Q_mX_0\rangle\langle Q_mX_0\vert \otimes\bigotimes_{l=k+2}^\infty
a_{00}^l.
\end{eqnarray*}
Hence, we get $$\mathrm{Tr}_{T\Phi}\left[\overline{P}^{k+1}_m
\mathcal{L}_{k+1}(\mathcal{P}\rho_k)\right]=\bigg\vert \sum_i
\langle Q_mX_0,X_i\rangle
U_{i0}\psi_k\bigg\rangle\bigg\langle\sum_j \langle
Q_mX_0,X_j\rangle
U_{j0}\psi_k\bigg\vert=\vert\theta_{k+1}\rangle\langle\theta_{k+1}\vert.$$
The normalisation factor $p_m^{k+1}$ gives the right expression of $\psi_{k+1}$ and
the recursion holds.
\end{pf}
\subsubsection{Non Markovian unravelling}

In this section we investigate the unravelling in the context of non-Markovian
quantum repeated measurements.  

Our discussion will be based on a typical physical
application, namely a two level atom (a qubit) in contact with a chain of spins. 
Mathematically, for this model $\mathcal{H}_0=\mathcal{H}=\b{C}^2$. Let us make precise the
description of the interaction. Let $k$ be fixed,
the interaction number $k$ is defined by a total Hamiltonian
$H_{\mathrm{tot}}(k)$ of the following form:
\begin{equation}\label{total}H_{\mathrm{tot}}(k)=H_0+H_{R}(k)+\lambda H_{I}(k),\end{equation}
where:
\begin{itemize} 
\item The operator $H_0$ corresponds to the
free Hamiltonian of $\mathcal{H}_0$. It is a self-adjoint operator
acting non-trivially on $\mathcal{H}_0$ and acting like the
identity operator on the chain.
 \item The operator $H_{R}(k)$ is the free
Hamiltonian corresponding to the free evolution of the $k$ first
copies of $\mathcal{H}$. It acts non-trivially on
$\bigotimes_{j=1}^k\mathcal{H}_j$ and like the identity operator
elsewhere. If $H$ corresponds to the free evolution of one copy of
$\mathcal{H}$, we have
$$H_{R}(k)=\sum_{i=1}^{k}\left(I\otimes\bigotimes_{j=1}^{i-1}I\otimes
H\otimes\bigotimes_{j>i+1}I\right)=\sum_{i=1}^{k}H^{(k)}.$$ 
In our context, the natural Hamiltonian of a single copy is $H=\gamma a_{00}$, where $\gamma$ is a real constant.
\item The self-adjoint operator $H_{I}(k)$ is 
the interaction Hamiltonian. It acts non-trivially on
$\mathcal{H}_0\otimes\bigotimes_{j=1}^k\mathcal{H}_j$ and like the
identity operator elsewhere.
The interactionHamiltonian $H_{I}(k)$ is given by
$$H_{I}(k)=\sum_{i=1}^k\left(C_i(k)a_{10}^{(i)}+C^\star_i(k)a_{01}^{(i)}\right),$$
where $C_i$ are operators on $\mathcal{H}_0$. In other words, each copy interacts with $\mathcal{H}_0$, but they do
not interact with each other. This is a simple physical case describing a non-Markovian
memory effect (at each new interaction, the previous copies continue
to interact).
 \item The term $\lambda$ is a real scalar corresponding to a coupling constant.
\end{itemize}

If the time of each interaction  is
$\tau$, then the unitary operator $U_k$ is defined as
$$U_k=e^{i\tau H_{tot}(k)}.$$
Hence, it describes a non-Markovian quantum repeated interactions.
The non-Markovian character comes from the definition of
$H_{I}(k)$.

\textbf{Remark:} It is worth noticing, that if $H_{I}(k)$ acts
non-trivially on the tensor product of $\mathcal{H}_0$ with
$\mathcal{H}_k$ and like the identity operator elsewhere, we
recover a Markovian approach of quantum repeated interactions. Actually, it does not correspond directly to the previous definition of Markovian interaction but it is equivalent to it. Indeed, at the $k$-th interaction the fact that $H_{I}(k)$ acts
non-trivially on the tensor product of $\mathcal{H}_0$ with
$\mathcal{H}_k$ and like the identity operator elsewhere, means that
only $C_k(k)$ is a non zero operator. Hence, the $(k-1)$ first copies do not interact with $\mathcal{H}_0$, while we can keep the definition of their free evolution. In the first definition of the Markovian case we do not refer to the free evolution of the $(k-1)$ first copies which could mean that we keep a kind of memory, but by taking the partial trace this memory disappears.
\bigskip

Our strategy to study the unravelling question in this context is the following. We start by considering the two first interactions, that is $\mathcal{H}_0$ coupled with $\mathcal{H}_1$ and next with $\mathcal{H}_1\otimes\mathcal{H}_2$. With a general approach, we show that the condition of unravelling imposes assumptions on the definition of the interaction. Next, we translate this assumptions in terms of the interaction model (\ref{total}).  Finally, by considering the small interaction time $\tau$, we show that unravelling imposes to "forget" the interaction with $\mathcal{H}_1$ during the second interaction, except for a very special case. Let us stress that forgetting the interaction with $\mathcal{H}_1$ during the second interaction corresponds to the Markovian assumption. 

The space describing the two first interactions is $\mathcal{H}_0\otimes\mathcal{H}_1\otimes\mathcal{H}_2$. Let $$U_1=\sum_{i,j=0,1}U_{i,j}\,a_{ij}^{1}\otimes I$$ be a unitary operator describing the first interaction and let $$U_2=\sum_{i,j=0,1}\sum_{k,l=0,1}U_{i,j,k,l}\,\,a_{ij}^{1}\otimes a_{kl}^{2}$$ be a unitary describing the second interaction. Let $A=\lambda_0P_0+\lambda_1P_1$ be the measured observable and let $\rho=\vert\psi\rangle\langle\psi\vert$ be the initial state of $\mathcal{H}_0$.

Concerning the observable $A$ there exist unitary matrices $Q_m$, $m=0,1$ such that $P_m=\vert Q_mX_0\rangle\langle Q_mX_0\vert$.  According to the unravelling result in the Markovian case, the first random state $\rho_1$ on $\mathcal{H}_0\otimes\mathcal{H}_1\otimes\mathcal{H}_2$ can take one of the values
$$\frac{1}{Z_1(m)}\vert \psi_1(m)\rangle\langle\psi_1(m)\vert\otimes \vert Q_mX_0\rangle\langle Q_m X_0\vert\otimes \vert X_0\rangle\langle X_0\vert,\,\,m=0,1$$
where $\psi_1(m)=\vert \sum_i \langle Q_mX_0,X_i\rangle
U_{i0}\psi\rangle$ and $Z_1$ corresponds to the normalisation factor. Hence the state $\mathcal{P}\rho_1$ is a pure state.
For the second interaction and the second measurement, we introduce the following notation
$$\mathcal{L}_{u0}=\Bigg(\sum_{i,j=0,1}U_{iju0}\,\,a_{ij}^{1}\Bigg)\otimes a_{u0}^{2},\,\,u=0,1.$$
In the same way, the second state $\rho_2$ can take one of the values
\begin{eqnarray*}&&\rho_2(m,n)=\\
&&\frac{1}{Z_2(m,n)}\Bigg(\sum_{u,v}\langle Q_nX_0, X_u\rangle\langle X_v,Q_nX_0\rangle\mathcal{L}_{u0}\bigg(\psi_1(m)\otimes\vert Q_mX_0\rangle\langle Q_mX_0\vert\bigg)\mathcal{L}_{v0}^\star\Bigg)\otimes\vert Q_nX_0\rangle\langle Q_nX_0\vert,
\end{eqnarray*}
with $(m,n)\in\{0,1\}^2$. The accurate computation of $\mathcal{P}\rho_2(m,n)$ gives
\begin{eqnarray*}\hspace{-2,5cm}\mathcal{P}\rho_2(m,n)=\frac{1}{Z_2(m,n)}\Bigg(\bigg\vert\mathcal{H}_0(m,n)\psi_1(m)\bigg\rangle\bigg\langle\mathcal{H}_0(m,n)\psi_1(m)\bigg\vert\\\hspace{2cm}+\bigg\vert\mathcal{H}_1(m,n)\psi_1(m)\bigg\rangle\bigg\langle\mathcal{H}_1(m,n)\psi_1(m)\bigg\vert\Bigg)\otimes a_{00}^{1}\otimes a_{00}^{2},
\end{eqnarray*}
where
$$\mathcal{H}_i(m,n)=\sum_{u,v}\langle Q_nX_0, X_u\rangle\langle X_v,Q_mX_0\rangle\,U_{i,u,v,0},\,\,i=0,1.$$
As a consequence, unravelling is possible if the following condition is satisfied
\begin{eqnarray*}&&\bigg\vert\mathcal{H}_0(m,n)\psi_1(m)\bigg\rangle\bigg\langle\mathcal{H}_0(m,n)\psi_1(m)\bigg\vert+\bigg\vert\mathcal{H}_1(m,n)\psi_1(m)\bigg\rangle\bigg\langle\mathcal{H}_1(m,n)\psi_1(m)\bigg\vert\\&=&\vert\psi_2(m,n)\rangle\langle\psi_2(m,n)\vert,
\end{eqnarray*}
for some vector $\psi_2(m,n)$.  If we want a rank one projector to be expressed as the sum of two rank one projectors there exist constants $\mu$ and $\nu$ such that
$$\nu\,\,\mathcal{H}_0(m,n)\psi_1+\mu\,\,\mathcal{H}_1(m,n)\psi_1=0.$$
In general, $\mu$ and $\nu$ will depend on $n$, $m$ and also on $\psi_1$. Here, we assume that there exist non zero constants  $\mu(m,n)$ and $\nu(m,n)$ such that
\begin{equation}\label{cond}\nu(m,n)\,\,\mathcal{H}_0(m,n)+\mu(m,n)\,\,\mathcal{H}_1(m,n)=0.\end{equation}
This assumption will be justified later after the discussion regarding unravelling.

Now, our aim is to show how the condition $(\ref{cond})$ is connected to the possibility of obtaining unravelling in the case of the interaction $(\ref{total})$. The condition $(\ref{cond})$ imposes a relation between the coefficients $U_{i,j,k,l}$ of the unitary operator $U_2$. Let us express these coefficients in the interaction model $(\ref{total})$. The total Hamiltonian for two interactions is defined by
\begin{eqnarray*}H_{tot}&=&H_0\otimes I\otimes I+I\otimes\gamma a_{00}\otimes I+I\otimes I\otimes\gamma a_{00}\\
&&+\lambda(C_1(2)\otimes a_{10}\otimes I+C_1^\star(2)\otimes a_{01}\otimes I\\&&\hspace{1cm}+C_2(2)\otimes I\otimes a_{10}+C_2^\star(2)\otimes I\otimes a_{01}).
\end{eqnarray*}
In order to describe the coefficients of $U_2$ in terms of $H_{\mathrm{tot}}$, we introduce an appropriate basis of $\mathcal{H}_0\otimes\mathcal{H}_1\otimes\mathcal{H}_2$, which is $X_0\otimes X_0\otimes X_0,\, X_1\otimes X_0\otimes X_0,\, X_0\otimes X_1\otimes X_0,\, X_1\otimes X_1\otimes X_0,\, X_0\otimes X_0\otimes X_1,\, X_1\otimes X_0\otimes X_1,\, X_0\otimes X_1\otimes X_1,\, X_1\otimes X_1\otimes X_1$. Hence, we can write 
\begin{equation}
H_{\mathrm{tot}}=\left(\begin{array}{cccc}
H_0+2\gamma I& \lambda C_1(2)&\lambda C_2(2)&0\\
\lambda C_1^\star(2)&H_0+\gamma I&0&\lambda C_2(2)\\
\lambda C_2^\star(2)&0&H_0+\gamma I&\lambda C_1(2)\\
0&\lambda C_2^\star(2)&\lambda C_1^\star(2)&H_0
\end{array}\right),\,\,\,U_2=\left(\begin{array}{cccc}
U_{0,0,0,0}&U_{0,1,0,0}&*&*\\U_{1,0,0,0}&U_{1,1,0,0}&*&*\\U_{0,0,1,0}&U_{0,1,1,0}&*&*\\
U_{1,0,1,0}&U_{1,1,1,0}&*&*
\end{array}\right).
\end{equation}
By studying the asymptotic expansion of $U_2=e^{i\tau H_{tot}}$ in term of $\tau$, one can find operators $L_{i,j,,k,l}(\tau)$ such that
\begin{eqnarray}\label{asympt}
U_{0,0,0,0}&=&I+\tau L_{000}(\tau),\,\,\,\,\,\,\,\,\,\,\,\,\,U_{1,1,0,0}=I+\tau L_{1100}(\tau),\\
U_{0,1,0,0}&=&\tau C_{1}(2)L_{0,1,0,0}(\tau),\,\,\,\,U_{1,0,0,0}=\tau C^\star_{1}(2)L_{1,0,0,0}(\tau),\\
U_{0,0,1,0}&=&\tau C^\star_{2}(2)L_{1,0,0,0}(\tau),\,\,\,\,
U_{1,1,1,0}=\tau C_{2}(2)L_{1,0,0,0}(\tau ),\\
U_{0,1,0,1}&=&\tau^2C_{1}(2)L_{1,0,0,0}(\tau),\,\,\,U_{1,0,1,0}=\tau^2C^\star_{1}(2)L_{1,0,0,0}(\tau).\label{asympt1}
\end{eqnarray}
Furthermore, each operator $L_{i,j,k,l}(\tau)$ converges to a non zero operator when $\tau$ goes to zero. 

 Now, by comparing this asymptotic description in terms of $\tau$ and the condition $(\ref{cond})$, we investigate the question of unravelling. The result depends on the form of the observable $A$. There are two different situations.
\bigskip

The first situation corresponds to the case where $P_0=\vert X_0\rangle\langle X_0\vert$ (this is equivalent to the case where $P_1=\vert X_0\rangle\langle X_0\vert$). In this case we have the corresponding $Q_0=I$, then $\langle X_0,Q_0X_o\rangle=1$ and the condition $(\ref{cond})$ for $m=n=0$ gives
$$\nu(0,0)U_{0,0,0,0}=\mu(0,0)U_{1,0,0,0}.$$
Hence, the asymptotic conditions $(\ref{asympt}-\ref{asympt1})$ impose $C_{1}(2)=0$. This situation corresponds to the Markovian case.

As a consequence, for any observable where $P_0=\vert X_0\rangle\langle X_0\vert$ (or $P_1$), the unravelling condition imposes the Markovian approach since the memory of the interaction with the first copy disappears Indeed, by recursion, for the third interaction and measurement, a similar reasoning shows that the unravelling condition imposes to forget the memory of the second copy and so on. 
\bigskip

The second situation concerns the case where $0<\langle X_0,Q_m X_0\rangle<1,\,\,m=0,1$. Indeed, if for $m\in\{0,1\}$, we have $\langle X_0,Q_m X_0\rangle\in\{0,1\}$, which corresponds to the first case. Now, the condition $(\ref{cond})$ and the asymptotic conditions $(\ref{asympt}-\ref{asympt1})$ give
$$\bigg(\nu(m,n)\langle Q_nX_0, X_0\rangle\langle X_0,Q_mX_0\rangle-\mu(m,n)\langle Q_nX_0, X_0\rangle\langle X_1,Q_mX_0\rangle\bigg)I=\tau\,W(\tau),$$
where $W(\tau)$ converges to an operator when $\tau$ goes to zero. Hence, it imposes that
\begin{equation}\label{cond1}
\frac{\nu(m,n)}{\mu(m,n)}=\frac{\langle X_1,Q_mX_0\rangle}{\langle X_0,Q_mX_0\rangle}.
\end{equation}
Now, let deal with the terms in $\tau^2$. If $C_1(2)\neq C_1(2)^\star$, we get an equality of the following form
$$\tau^2Z(\tau)=\tau Y(\tau),$$
where $Z(\tau)$ and $Y(\tau)$ converge to non zero operators, when $\tau$ goes to zero. As a consequence, if $C_1(2)\neq C_1(2)^\star$, we cannot get unravelling.

Let us now suppose that $C_1(2)=C_1(2)^\star$. The condition $(\ref{cond})$ and the asymptotic conditions $(\ref{asympt}-\ref{asympt1})$  give 
\begin{eqnarray*}&&\tau^2\bigg(\nu(m,n)\langle Q_nX_0, X_1\rangle\langle X_1,Q_mX_0\rangle-\mu(m,n)\langle Q_nX_0, X_1\rangle\langle X_0,Q_mX_0\rangle\bigg)C_1(2)N(\tau)\\&&=\tau\,M(\tau),\end{eqnarray*}
where $N(\tau)$ and $M(\tau)$ converge to non zero operators when $\tau$ goes to zero. It imposes the condition
\begin{equation}\label{cond2}
\frac{\nu(m,n)}{\mu(m,n)}=\frac{\langle X_0,Q_mX_0\rangle}{\langle X_1,Q_mX_0\rangle},\,\,\textrm{or}\,\,C_1(2)=0.
\end{equation}
The case $C_1(2)=0$ corresponds to the Markovian approach and the result in Section $\ref{tt}$ imposes naturally the unravelling result. Otherwise, the condition $(\ref{cond1})$ combined with the condition $(\ref{cond2})$ implies that
$$\langle X_0,Q_mX_0\rangle^2=\langle X_1,Q_mX_0\rangle^2,\,\,m=0,1.$$
It is easy to notice that, this condition corresponds to special observables of the form
\begin{equation}\label{form}
A=\lambda_0\,\left(\begin{array}{cc} 1/2&1/2\\1/2&1/2\end{array}\right)+\lambda_1\,\left(\begin{array}{cc} 1/2&-1/2\\-1/2&1/2\end{array}\right).
\end{equation}
This means that in the non-Markovian case with the interaction model $(\ref{total})$, the unravelling result can not be obtained if the measured observable is not of the form  $(\ref{form})$ with $C_1(2)=C_1(2)^\star$.

Reciprocally, we can now exhibit a special model of interaction $(\ref{total})$ that allows unravelling. We consider that for all $(i,k)\in\b{N}$, we have $C_i(k)=C_i(k)^\star$ (this corresponds to the assumption $C_1(2)=C_1(2)^\star$). Furthermore, we assume that there is not free evolution concerning the infinite chain, that is $\gamma=0$. Moreover, to use directly the previous discussion, we suppose that at each interaction only two copies of $\mathcal{H}$ interact with $\mathcal{H}_0$. Mathematically, for $k\geq2$, at the $k$-th interaction, we consider that only the copies number $k$ and number $k-1$ interact with $\mathcal{H}_0$. In other words, for $k\geq 2$, at the $k$-th interaction, we have $C_i(k)=0$ for all $1\leq i\leq k-2$. Now, for an observable of the form $(\ref{form})$, by computing $e^{i\tau H_{\mathrm{tot}}}$, it is easy to see that
$$\mathcal{H}_0(m,n)=\mathcal{H}_1(m,n),\,\,m,n=0,1$$
and the condition (\ref{cond}) is fulfilled. 
As only the last incoming copy and the previous one interact with $\mathcal{H}_0$ at each interaction, a recursive reasoning shows that we obtain unravelling in this case.
\bigskip

Let us finish by justifying that the condition
$$\nu\,\,\mathcal{H}_0(m,n)\psi_1+\mu\,\,\mathcal{H}_1(m,n)\psi_1=0,$$
for constant $\nu$ and $\mu$ depending on $n,m$ and $\psi_1$ implies that
$$\nu(m,n)\,\,\mathcal{H}_0(m,n)+\mu(m,n)\,\,\mathcal{H}_1(m,n)=0,$$
without dependence in $\psi_1$. This result stems from the following proposition in linear algebra.

\begin{pr}
Let $D_1=\{\psi\in\b{C}^2/\Vert\psi\Vert=1\}$ be the unit disk. Let $\phi$ be a linear application such that for all $\vert x\rangle\in D_1$ there exist $\lambda$ such that $\phi(\vert x\rangle)=\lambda\vert x\rangle$, then there exists $\lambda$ such that for all $\vert x\rangle$, we have $\phi(\vert x\rangle)=\lambda\vert x\rangle$, that is $\phi=\lambda I$.
\end{pr}

In our case, we aim to apply this result to $\mathcal{H}_0(m,n)^{-1}\mathcal{H}_1(m,n)$ for appropriate $m$ and $n$. Indeed, it is worth noticing that for $\tau$ small enough, there always exists $(m,n)\in\{0,1\}$ such that the operator $\mathcal{H}_0(m,n)$ is invertible. Indeed there always exists $(m,n)\in\{0,1\}$ such that $\mathcal{H}(m,n)$ is of the form $\alpha I+\tau F(\tau)$, where $F(\tau)$ converges when $\tau$ goes to zero and $\alpha$ is non null.\\
\textbf{Remark:} For all non diagonal observables this property is satisfied for all couples $(m,n)$. For the diagonal observable, we can consider the case $m=n=0$.\\
To conclude, it remains to prove that random vectors $\psi_1$ cover all the unit disk. This is justified as follows. The rules of the first interaction and the first measurement give rise to a random transformation $\mathbf{\Lambda}$ from $D_1$ to $D_1$, which maps $\psi_0$ in a random unit vector $\psi_1$.

 For all observable $A$, it is worth noticing that for a $\tau$ small enough there exists $m\in\{0,1\}$ such that the operator $\sum_{i}\langle Q_mX_0,X_i\rangle U_{i0}$ is invertible (for the non diagonal observable for all $m$ this property is satisfied, for the diagonal observable it corresponds to the case m=0). As we must consider all the possible results for the measurement, in any case, there exists at least one possibility that the range of $\mathbf{\Lambda}$ is $D_1$ and the result holds.

\end{document}